# Robust Superconductivity and High Upper Critical Fields in Epitaxial cubic W₂N Thin Films

Aditya Singh[1], Arnaud le Febvrier[2], Sanath Kumar Honnali[2], Abhisek Mishra[3], Grzegorz Greczynski[4], Subhankar Bedanta[3], Per Eklund[2, *], Ajay Soni[1, *]

[1]School of Physical Sciences, Indian Institute of Technology Mandi, Mandi 175005, Himachal Pradesh, India
[2]Department of Chemistry - Ångström Laboratory; Inorganic Chemistry, Uppsala University, Uppsala, 75105, Sweden
[3] Laboratory for Nanomagnetism and Magnetic Materials (LNMM), School of Physical Sciences, National Institute of Science Education and Research (NISER), An OCC of Homi Bhabha National Institute (HBNI), Jatni 752050 Odisha, India.
[4]Thin Film Physics Division, Department of Physics, Chemistry, and Biology (IFM), Linköping University, Linköping SE-581 83, Sweden
*Author to whom correspondence should be addressed: per.eklund@kemi.uu.se, ajay@iitmandi.ac.in

Transition Metal Nitrides (TMNs) are a versatile class of materials, combining chemical robustness, high hardness, and superconducting behaviour with critical temperatures between 2–10 K. While several binary TMNs have been explored, superconductivity in stoichiometric W₂N has remained largely unexplored. Here, we report on superconducting thin films of stoichiometric W₂N, demonstrating a distinctly high upper critical field ($H_{c2}$) of ~ 8.5 T, uncommon among binary TMNs. This robust superconducting response under high magnetic fields highlights the technological relevance of W₂N for integrated quantum and cryogenic electronic platforms. Overall, these results position stoichiometric W₂N as a promising addition to the TMN superconducting landscape, opening new avenues for functional materials design based on chemically stable and mechanically resilient nitrides.

**Introduction:**

Transition metal nitrides (TMNs) constitute a versatile class of materials characterized by a unique interplay of metallic and covalent bonding, which imparts an exceptional combination for mechanical robustness, thermal stability, and electronic tunability [1]. The bonding hybridization underlies their widespread use in applications such as diffusion barriers, protective coatings, cutting tools.[2-5] Beyond their mechanical resilience, TMNs exhibit rich electronic phenomena and applications, ranging from metal–insulator transitions [6], ferromagnetism [7], spintronic functionalities [8] and superconductivity[9]. The incorporation of nitrogen atoms strengthens the metal–nitrogen bonds and enhances electron–phonon (e-ph) coupling, which plays a pivotal role in mediating superconductivity.[10] Several TMNs such as InN, HfN, ZrN, Mo₂N, Nb₂N, ReN, RuN, TiN, and VN [11-18] have been explored for their low temperature electronic transport properties, underscoring their potential as tuneable superconducting platforms. Furthermore, TMNs can crystallize in several structural polymorphs, such as hexagonal,[19] cubic,[18] orthorhombic,[20] rhombohedral,[21] and tetragonal phases,[22] each associated with distinct electronic structures and phonon dynamics. The superconducting properties are strongly structure-dependent, for instance, hexagonal MoN exhibits a relatively high critical temperature ($T_c \sim 13$ K),[23] whereas cubic $Mo_2N$ typically shows a lower $T_c \sim 5–6$ K.[18] revealing the sensitivity of superconductivity to lattice symmetry and bonding environment. Nitrogen stoichiometry also plays a decisive role, as even small deviations can significantly alter carrier concentrations, induce lattice strain, and modulate e-ph coupling strength.[24]

Among TMNs, tungsten nitrides (W–N) are particularly interesting due to their polymorphic nature, and multiple applications such as mechanical hardness, diffusion barrier performance, catalytic activity, and electrochemical energy storage capabilities [25-29]. Multiple phases such as cubic $W_3N_4$, hexagonal and rhombohedral $W_2N_3$,[21] cubic WN,[30] and cubic $W_2N$, have been reported, each with distinct structural and electronic characteristics. However, despite this structural richness, the relationship between phase formation, nitrogen stoichiometry, and superconducting behaviour in the W–N system remains poorly understood. The physical properties of these materials are highly sensitive to the W:N ratio and crystallographic phase, thus controlling stoichiometry within a specific crystal structure is important for getting bespoke properties. Theoretical predictions[31] have suggested $W_2N$ as a potential superconductor coexisting with a possible charge-density-wave instability, yet experimental validation has been scarce. Prior studies have primarily focused on tungsten thin films synthesized under varying $N_2$: Ar atmospheres, yielding multiphase materials comprising metallic W, nitrogen-stabilized β-W, and $W_2N$ inclusions [32-34]. Consequently, the intrinsic superconducting properties of stoichiometric $W_2N$ have remained ambiguous. Previous reports have indicated superconductivity near 1.3 K, only marginally higher than that of pure tungsten ($T_c \sim 0.011$ K),[35] in sharp contrast to other TMNs where nitrogen incorporation substantially enhances $T_c$ for instance, in case of Nb, $T_c \sim 9$ K, while NbN has $T_c \sim 18$ K.[36, 37] Thus, in case of $W_2N$, the role of structure, stoichiometry, and bonding provides a platform to explore the superconductivity in this material.

In this work, we address this research gap by presenting the first comprehensive experimental investigation of superconductivity in stoichiometric $W_2N$ thin films. The films are grown via reactive DC sputtering on $Al_2O_3$ and MgO substrates with an average thickness of $\sim 270$ nm, exhibiting a clear superconducting $T_c$ around $\sim 5K$. The superconducting behaviour is further corroborated by a distinct diamagnetic transition observed within the same temperature range, confirming the bulk nature of the superconductivity. Magnetization and resistance measurements at different applied magnetic fields are employed to determine the lower ($H_{c1}$) and upper critical fields ($H_{c2}$), revealing values in the range of $H_{c1} \sim 50$ Oe and $H_{c2} \sim 8$ T respectively. This comparatively high upper critical field, coupled with a relatively large intermediate state, indicates that $W_2N$ shows strong type-II superconductivity. This characteristic is particularly significant for potential technological applications, such as in the development of high-performance single-photon detectors and other superconducting devices.

**Results and discussion:**

The optimized growth parameters were determined through preliminary depositions on relatively thinner films of 54 nm after confirmation from XRD and elemental composition by XPS. The XPS analysis confirmed a near-stoichiometric composition with an N/W ratio of 0.52 ($W_2N_{1.04}$), while XRR analysis provided a deposition rate of 1.8 nm/min. Thicker films of 270 nm are deposited under the same condition on the same day, presenting the same crystal and composition characteristics as the thinner film used during the optimization process (Figure S1).

The composition of the thick films is confirmed using a combination of RBS ToF-ERDA, which together allow accurate quantification of both the heavy element tungsten sublattice and the lighter element N despite their strong mass contrast. Both films are confirmed to be tungsten nitride ($W_2N$) with compositions very close to the ideal 2:1 ratio. Specifically, $W_2N/Al_2O_3$ exhibits a composition of $W_2N_{1.08}$, whereas $W_2N/MgO$ gives $W_2N_{1.14}$. Such small deviations from exact stoichiometry are well within the tolerance range typically reported for sputtered transition-metal nitrides and are consistent with stable $W_2N$ phase formation. All the composition results are presented in Table S1.

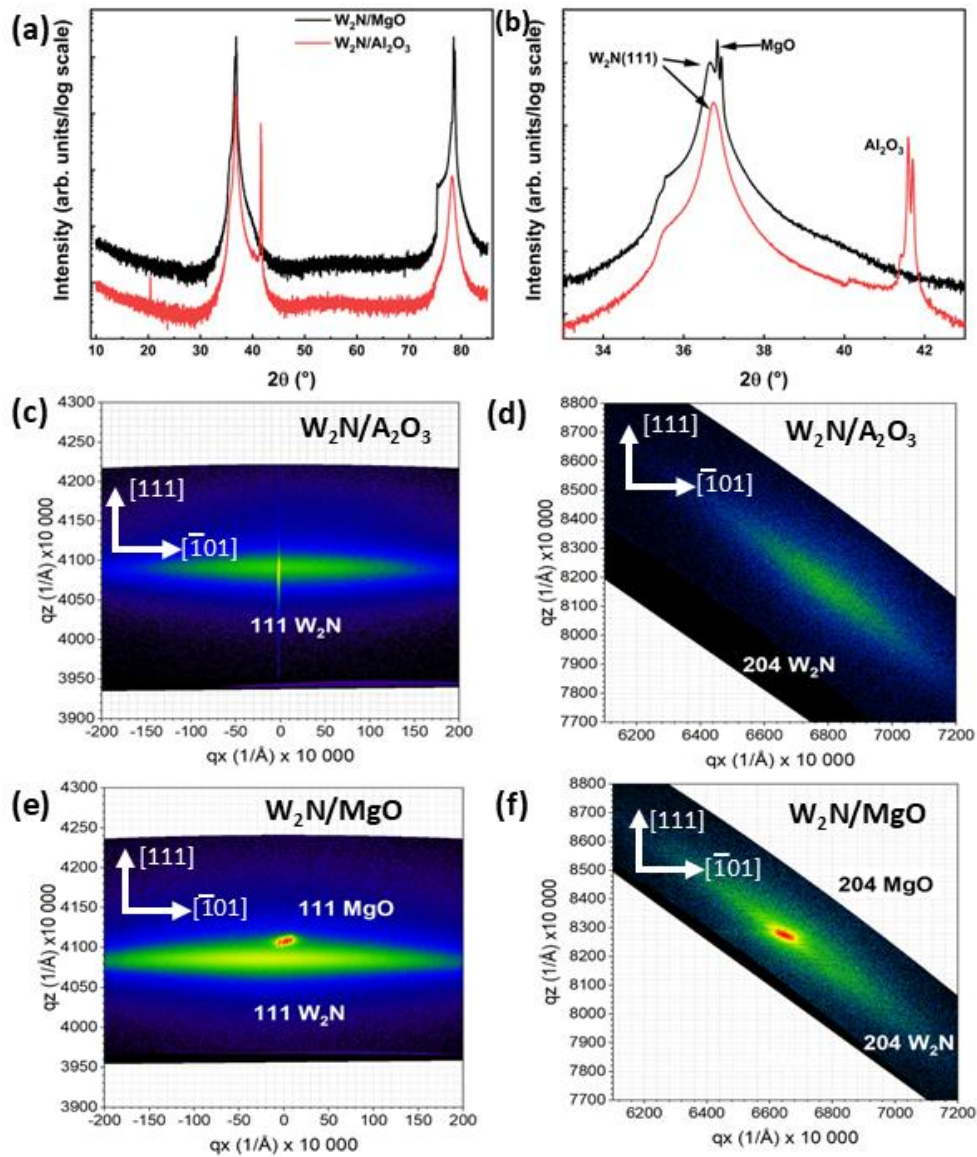

Figure 1: XRD patterns of $W_2N/MgO$ and $W_2N/Al_2O_3$ in (a)10° - 85° and (b) 33°- 43°. RSM of (c, d) $W_2N/Al_2O_3$ and (e, f) $W_2N/MgO$ films around the symmetric 111 and asymmetric 204 reflections, respectively.

Figure. 1(a) shows the full-range XRD 2θ scan for both $W_2N/Al_2O_3$ and $W_2N/MgO$, confirming the overall phase purity of the films without any detectable secondary phases across the scanned range. Figure. 1(b) presents the the magnified 2θ window around 34–42° where the dominant film and substrate peaks appear more clearly. The substrate reflections are identified at 2θ ~ 41.6° and at 2θ ~ 36.8° identified at 0001 $Al_2O_3$ (PDF 00-046-1212) and 111 MgO (PDF 01-071-6452). The diffraction peak observed at 2θ ~ 36.7° (d-spacing = 2.447 Å) for $W_2N/Al_2O_3$ and at 2θ ~ 36.6° (d-spacing 2.453 Å) for $W_2N/MgO$ is indexed to the 111 reflections of cubic $W_2N$ (PDF 00-025-1257), indicating a preferential out-of-plane (111) orientation. Pole Figure analysis (Figure. S2) is used to determine the epitaxial relationship for both the films which are as follows: (111) $W_2N$ ∥ (0001) $Al_2O_3$ (out-of-plane) and [11$\bar{2}$]$W_2N$ ∥ [10$\bar{1}$0] $Al_2O_3$ (in-plane); and (111) $W_2N$ ∥ (111) MgO (out-of-plane) and [110] $W_2N$ ∥ [110] MgO (in plane).

In order to evaluate epitaxy, lattice coherence, and strain state of the films RSM is performed. The Figure. 1(c-f) display the symmetrical RSM around the 111 reflections of the film and the asymmetric RSM taken around the 204 peak from the film for both films. Note that the out of plane direction ($q_z$) correspond to the [111] direction while the in-plane direction ($q_x$) corresponds to the [$\bar{1}$01] of $W_2N$. The symmetrical RSM confirm the Bragg Brentano measurement while the asymmetrical RSM confirm the epitaxial character of the film with the possibility to resolve the structure of $W_2N$. On c-plane sapphire (Figure. 1(c, d)) the extracted lattice constants (a = 4.188 Å, $\alpha$ =90.62°) reveals an in-plane strain of +0.52% and out-of-plane compressive strain of –1.09%. On MgO (111) (Figure 1 (e, f)), sharp reflections from both the substrate and film appear on both RSM rendering challenges into the analyses but also confirmed the small deviation from a cubic material for $W_2N$ when compared to the substrate. The extracted lattice parameters (a = 4.190 Å, $\alpha$ = 90.67°) reveal that the film has an in-plane strain of +0.58% and a compensating out-of-plane compressive strain of –1.19%. On both substrates, $W_2N$ prosses relatively the same structure, with the same lattice parameters, thus expected to be same stress value. Comprehensive structural, compositional, and crystallographic characterization confirms that the synthesized thin films are single phase $W_2N$ with NaCl like structure with half-filled nitrogen sites, providing a solid foundation for the investigation of their low-temperature transport properties.

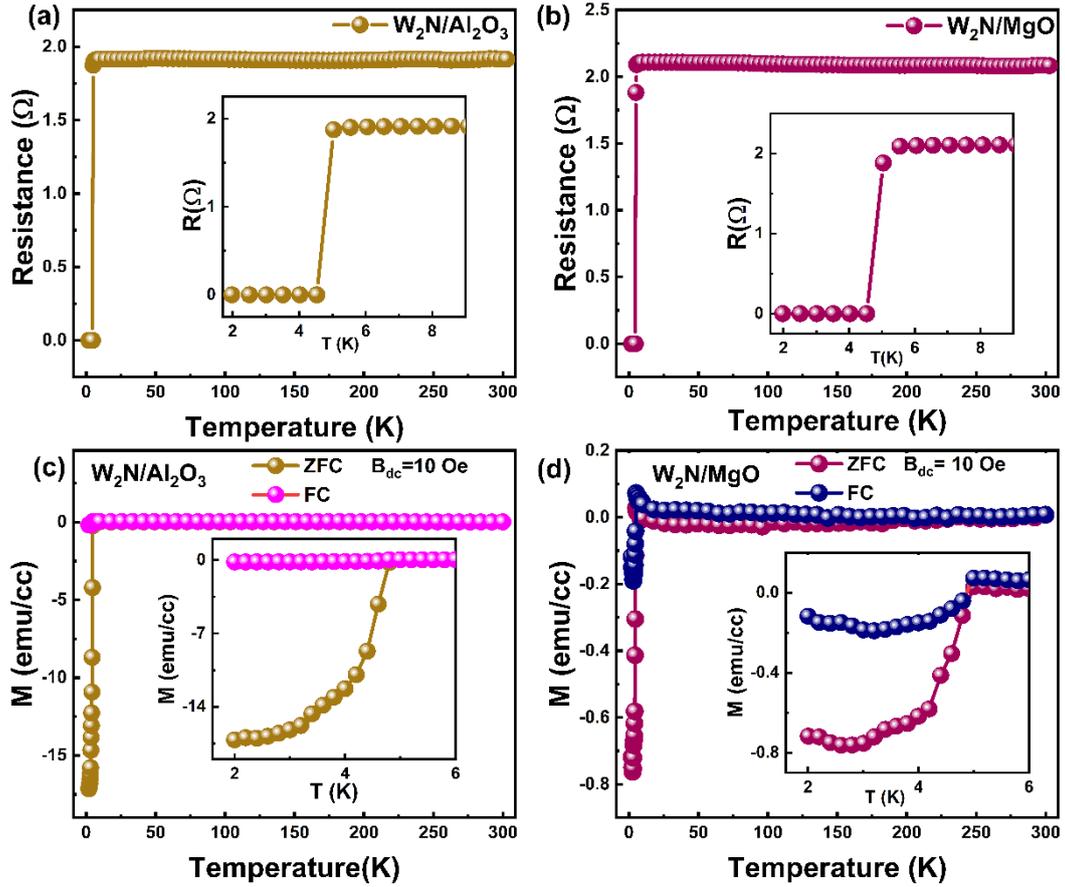

Figure 2: Temperature dependent Resistance (R(T)) of (a) $W_2N/Al_2O_3$ and (b) $W_2N/MgO$, ZFC and FC M-T curves for (c) $W_2N/Al_2O_3$ and (d) $W_2N/MgO$, (insets: low-temperature data).

Figure. 2(a-b) shows the temperature dependent resistance plots of the thin films. The critical temperature ($T_c$) for both the films has been evaluated using the 90% criteria and found out to be ~ 5 K for $W_2N/Al_2O_3$ and ~ 5.06 K for $W_2N/MgO$. In order to get a understanding about the underlying conduction mechanism in $W_2N$ thin films, the resistance data is analysed with small polaron hopping (SPH) model which relates the temperature dependent resistance $R(T)$ by the given expression $\rho(T) = \rho_0 T exp\left(\frac{E_p}{k_b T}\right)$, where $E_p$ is the polaron activation energy. A good fit to the SPH model (Figure S3 in Supplementary Information), indicates that electrical transport is dominated by polaron hopping arising from strong electron-phonon interactions, rather than electron transport.[38] Further onset of superconductivity is corroborated by the DC magnetization measurements (Figure 2(c, d)) with both the films showing sharp diamagnetic transition around ~5 K.

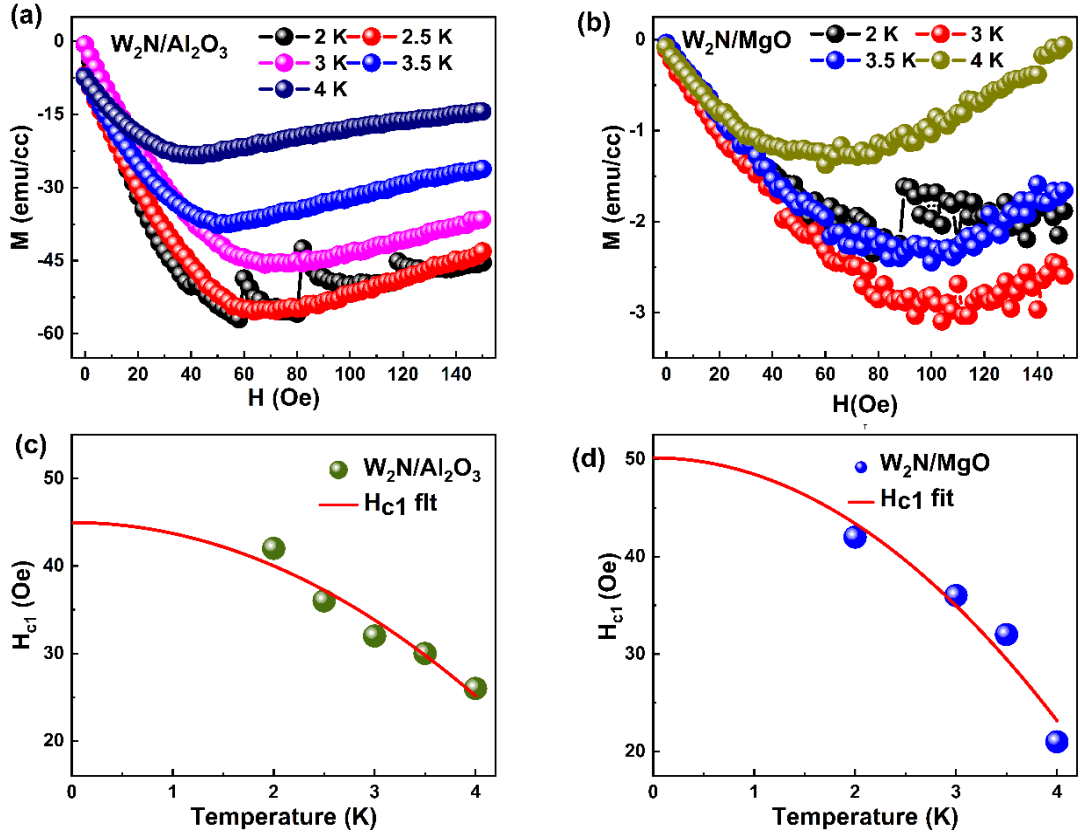

Figure. 3: M- H plots (a, b) and $H_{c1}$ versus T (c, d) plots of $W_2N/Al_2O_3$ and $W_2N/MgO$, respectively. The red line denotes the fitting using model provided in the text.

The lower critical field ($H_{c1}(T)$) at different temperatures, have been estimated from the M-H measurements performed below $T_c$ by using the Meissner–Ochsenfeld criterion.[18] Furthermore, to extract the $H_{c1}(0)$ the $H_{c1}$ vs $T$ data has been fitted using the equation $H_{c1} = H_{c1}(0)[1 - \left(\frac{T}{T_c}\right)^2]$. The obtained values of $H_{c1}(0)$ are found to be ~ 45 Oe for $W_2N/Al_2O_3$ and ~ 50 Oe for $W_2N/MgO$.

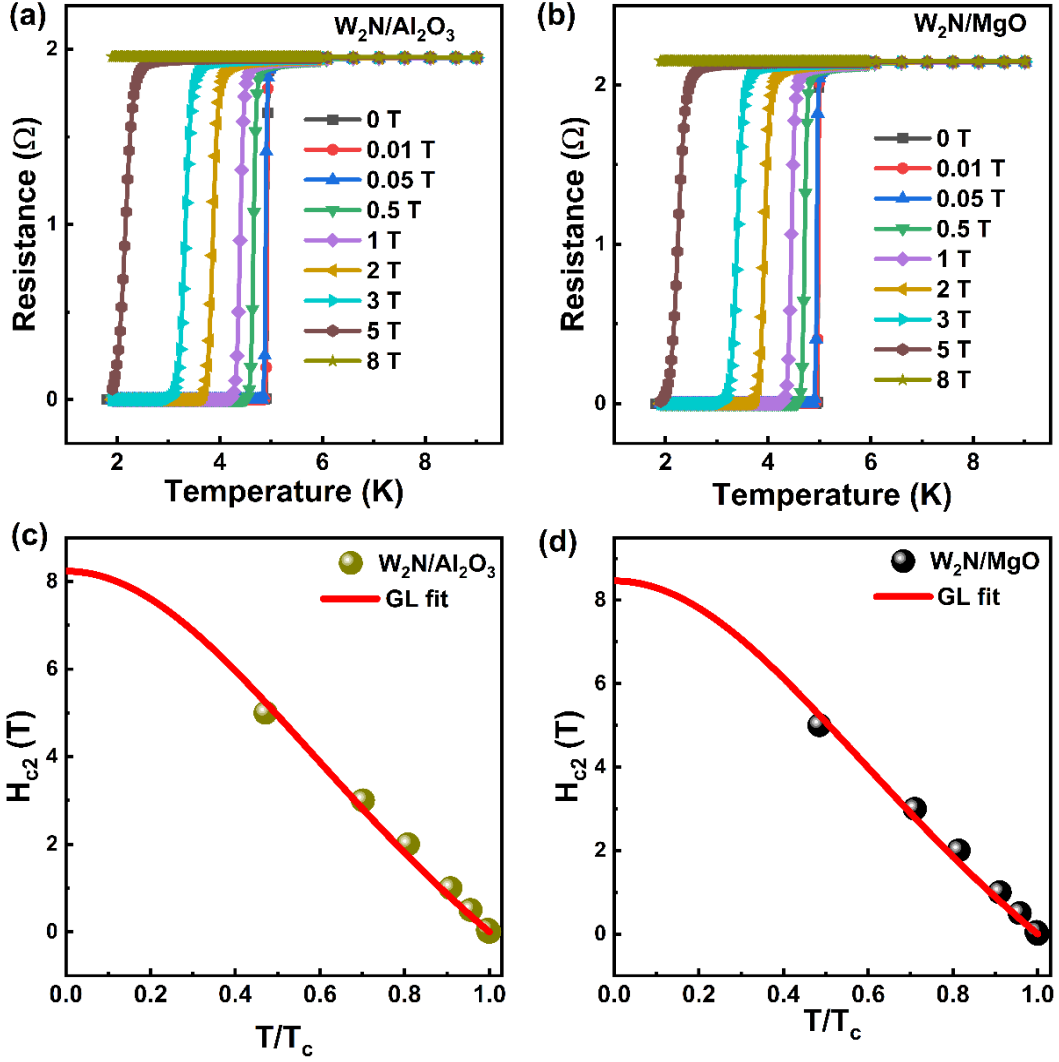

Figure 4: Low temperature resistance at different applied magnetic field for (a)W₂N/Al₂O₃, (b) W₂N/MgO. *Hc2* vs *T/Tc* for (c) W₂N/Al₂O₃, (d) W₂N/MgO. The red line denotes the fitting of *Hc2* with GL model.[18]

The $H_{c2}$ is a crucial metric determining the operational range of superconductors in high-field applications. Low temperature magnetoresistance data reveal the shifting of critical temperature to lower side, a characteristic feature of type II superconductors. $H_{c2}$ has been estimated from the $T_c^{onset}$ criteria where the normal state resistance value which is taken at 6 K. The obtained $H_{c2}$ vs $T$ has been fitted using the Ginzburg Landau (GL) relation  to obtain $H_{c2}(0)$, given by $H_{c2}(T) = H_{c2}(0)\frac{(1-t^2)}{(1+t^2)}$, where $t$ is the reduced temperature given by $t = \frac{T}{T_c}$ and $H_{c2}(0)$ is critical field at 0 K. $H_{c2}(0)$ estimated from GL model is found to be ~ 8.23 T and ~ 8.46 T for W₂N/Al₂O₃ and W₂N/MgO respectively. The estimated H$_{c2}$ values for W₂N rank among the highest reported for transition metal nitride superconductors[11-18], highlighting the exceptional high-field robustness of these epitaxial films. The *Hc2(0)* is related to the normal state resistivity ($\rho_n$) by the relation given by WHH approximation given by $H_{c2}(0) = 0.69\, T_c \frac{4eck_b}{\pi} N(0)\rho_n$, where $e, c, k_b, N(0)$ corresponds to electronic charge, speed of light, Boltzmann constant, density of states at Fermi level respectively.[17] The high *Hc2(0)* in $W_2N$ can be ascribed to a combination of factors such as high $T_c$, higher $N(0)$ and high $\rho_n$ as compared to the similar family of compounds.[11-18] Further, we have extracted the zero-temperature orbital limited upper critical field $H_{c2}^{orb}(0)$ from the Werthamer-Helfand-Hohenberg (WHH) formula given as $H_{c2}(0) = -0.693 T_c (\frac{dH_{c2}}{dT})_{T=T_c}$.[18] The calculated value of $H_{c2}(0)$ is 6.67 T for $W_2N/Al_2O_3$ and 6.82 T for MgO-111. The obtained $H_{c2}(0)$ from WHH approximation is lower than the ones obtained from GL fitting. Both these critical fields are less than Pauli Paramagnetic limit which is given by $H_p = 1.86 T_c$ (9.3 T for $W_2N/Al_2O_3$ and 9.41 T for $W_2N/MgO$ and therefore indicating of a conventional BCS type superconductivity. In order to gain a deeper insight about the mechanisms governing the upper critical field we have calculated the Maki parameter ($\alpha$), which is given by $\alpha = \sqrt{2} \frac{H_{c2}^{orb}(0)}{H_p(0)}$ . [18]In general, superconductivity can be destroyed either by the orbital pair-breaking mechanism, where the Lorentz force on Cooper pairs suppresses phase coherence, or by the Pauli paramagnetic effect, in which the Zeeman splitting of electronic spins breaks the singlet pairing. The competition between these two mechanisms determines the limiting behaviour of $H_{c2}$, with the Pauli limit becoming increasingly important in clean superconductors with short coherence lengths and large g-factors. The obtained values of $\alpha$ are ~1.01, 1.02 for $W_2N/Al_2O_3$ and $W_2N/MgO$ respectively. The observed Maki parameter, α, exceeding 1 signifies that the Pauli paramagnetic limiting effect is substantial and comparable to, or even dominant over, orbital pair-breaking.[39] Although α < 1.8, still greater than 1, thereby suggesting a potential formation of a Fulde-Ferrell-Larkin-Ovchinnikov (FFLO) superconducting state, where Cooper pairs exhibit a finite momentum due to the spin imbalance induced by the magnetic field.[40] The GL coherence length ($\xi_{GL}$), which characterizes the spatial scale over which the superconducting order parameter recovers in the presence of perturbations,given by $\xi_{GL} = \sqrt{\frac{\phi_0}{2\pi H_{c2}(0)}}$, where $\phi_0 = 2.068 \times 10^{-15}$ T. $m^2$ is magnetic flux quantum is calculated for both the films and found out to be ~ 6.32 nm for $W_2N/Al_2O_3$ and ~ 6.23 nm for $W_2N/MgO$. The GL London penetration depth ($\lambda_{GL}$) , which measures how far a magnetic field can penetrate into a superconductor before being screened by supercurrents, is evaluated from the equation given by $H_{c1} = \frac{\phi_0}{4\pi\lambda_{GL}^2} \ln\left(\frac{\lambda_{GL}}{\xi_{GL}}\right)$ and is found to be ~ 388 nm for $W_2N/Al_2O_3$ and ~ 366 nm for $W_2N/MgO$. Furthermore, we have also evaluated GL parameter ($\kappa_{GL}$) given by $\kappa_{GL} = \frac{\lambda_{GL}}{\xi_{GL}}$, a strong indicator of type of superconductivity presents in the system and found to be ~56 for $W_2N/Al_2O_3$ and ~59 for $W_2N/MgO$, significantly larger than $\frac{1}{\sqrt{2}}$, indicating a strong type-II superconductivity in all the synthesized thin films. The thermodynamical critical field ($H_c$) is evaluated using the relation $H_{c1}(0)H_{c2}(0) = H_c^2 \ln \kappa_{GL}$ which is found to be ~ 960 Oe and ~ 1020 Oe for $W_2N/Al_2O_3$ and $W_2N/MgO$ respectively. All the estimated parameters for the thin films are tabulated in Table 1.

Table 1: Thin Film characteristics of $W_2N$ on different substrates

| Sample | t (nm) | $T_c$ (K) | $H_{c1}(0)$ (Oe) | $H_{c2}(0)$ (T) | $H_{c2}^{orb}(0)$ (T) | $H_c$ (Oe) | $\xi_{GL}$ (nm) | $\lambda$ (nm) | $\alpha$ |
|---|---|---|---|---|---|---|---|---|---|
| $W_2N/Al_2O_3$ | 270 | 5 | 45 | 8.23 | 6.67 | 960 | 6.32 | 388 | 1.01 |
| $W_2N/MgO$ | 270 | 5.06 | 50 | 8.46 | 6.82 | 1020 | 6.23 | 366 | 1.02 |

**Conclusion:**

In summary, we demonstrate that stoichiometric $W_2N$ thin films exhibit robust type-II superconductivity with an exceptionally high upper critical field of ~ 8.5 T, positioning them among the strongest magnetic field resilient binary transition metal nitrides known to date. Analysis of the Maki parameter ($\alpha \sim 1$) indicates that Pauli paramagnetic effects are comparable to orbital pair-breaking, suggesting that $W_2N$ could serve as a promising platform for exploring exotic superconducting states such as the FFLO phase. The observed type-II superconductivity, combined with robust chemical stability and epitaxial quality, highlights the potential of $W_2N$ for both fundamental quantum materials research studies and the development of next-generation superconducting technologies. Overall, this work shows how precision thin-film synthesis can unlock emergent quantum behaviour in chemically resilient nitrides, broadening the design landscape for superconductors with tuneable electronic ground states.

**Experimental Section** – $W_2N$ thin films were grown on *c*-plane sapphire ($W_2N/Al_2O_3$) and MgO (111) ($W_2N/MgO$) substrates by reactive DC magnetron sputtering in an ultrahigh vacuum chamber.[41] Prior to deposition, the MgO substrate was cleaned sequentially with a Hellmanex soap solution (3 min), deionized (DI) water (5 min × 2), and then sonicated in acetone (10 min) and ethanol (10 min) to remove surface hydroxides and carbonates and were annealed at ~ 800° C for two hours.[42] After an optimization process the optimum deposition condition for obtaining $W_2N$ were a substrate temperature of ~ 200° C, deposition pressure of ~ 0.32 Pa with a mixture of $N_2$/Ar with a flow ratio of ~ 41 sccm/ 18 sccm 70% $N_2$ in the plasma. The base pressure of the chamber, at room temperature, was $1.3 \times 10^{-9}$ Torr whereas increased to $2.4 \times 10^{-9}$ Torr at 200° C. Before depositing the final thick $W_2N$ film, several preliminary depositions were carried out and systematically analysed using X Ray Diffraction (XRD) and X-Ray Photoelectron Spectroscopy (XPS) to achieve the desired stoichiometry and to optimize the deposition parameters for $W_2N$. XPS and XRR were performed on a thinner $W_2N$ film grown under identical deposition conditions as the samples presented in this manuscript to determine stoichiometry and deposition rate. The sample was transferred from the UHV deposition chamber to the XPS system with less than 3 minutes of atmospheric exposure to minimize contamination.

XPS analyzes of $W_2N$ films were conducted in an Axis Ultra DLD instrument from Kratos Analytical (UK) with the base pressure lower than $1.1\times10^{-9}$ Torr ($1.5\times10^{-7}$ Pa). Monochromatic Al Kα radiation (hν = 1486.6 eV) was used with the anode power set to ~ 150 Watt and the core level spectra W (*4f*) and N (*1s*) were recorded at normal emission angle. The analyzer pass energy was set at 20 eV, yielding a full width at half maximum of Ag $3d_{5/2}$ peak (~ 0.55 eV) of Ag foil as a reference. To ensure accurate N/W quantification, samples were analyzed in the as received state, thus avoiding preferential N re-sputtering by $Ar^+$ ions.[43] The analysis area was $0.3 \times 0.7$ mm$^2$. Spectrometer calibration was verified by measuring Au $4f_{7/2}$, Ag $3d_{5/2}$, and Cu $2p_{3/2}$ peak positions from sputter-etched Au, Ag, and Cu samples and comparison to the recommended ISO standards for monochromatic Al Kα sources.[44]

The elemental composition of the films was measured by combination of time-of-flight-Elastic Recoil Detection analysis (ToF-ERDA) and Rutherford Backscattering Spectrometry (RBS). The measurements were carried out using the Pelletron Tandem accelerator (5MV NEC-5SDH-2).[45] The ToF-ERDA measurements were conducted using 36 MeV primary iodine ion ($^{127}I^{8+}$) beam. The incident beam angle to the target surface normal was 67.5°, while the ToF-telescope and the gas ionization detector were placed at 45° relative to the incident beam direction. The depth profile of the elemental composition was acquired from the ToF-ERDA time and energy coincidence spectra using the Potku (version 2.2.4) code.[46] For RBS measurements, a 2 MeV He$^+$ primary beam was used, incident at 5° to sample surface normal and the backscattered ions were collected at 170° and the recorded spectra were analyzed using SIMNRA (version 7.03) software.[47]

X-ray diffraction (XRD) measurements were performed on a PANalytical X'Pert PRO diffractometer in Bragg–Brentano (θ–2θ) geometry with a Cu Kα (λ = 1.540598 Å) radiation and Ni filter. The recorded 2θ range is 10-90º with a step size of 0.008º and an equivalent time/step of 19 sec using the PIXcel 1D detector. Pole Figures were acquired with a polycapillary with a crossed slits (4 mm$^2$) as primary optics and parallel plate collimator (0.18°) (phi steps size = 1° and chi step size = 1°). Reciprocal Space Maps (RSM) of the symmetric 111 peak and asymmetric 204 peak of W$_2$N were acquired on the same type of diffractometer equipped with hybrid mirror as primary optics and a PiXcel2D detector for fast acquisition with no particular optics for the receiving optics. The XRR was performed using X-Ray mirror with 1/32° divergence slit in primary and parallel plate collimator in secondary optics.

The low temperature electrical transport (2-300 K) measurements were carried out using the physical property measurement system (PPMS, Quantum Design) equipped with a 14 T magnet. The temperature dependent magnetization measurements were carried out using an MPMS3 SQUID-VSM magnetometer (Quantum Design make) in Zero-Field-Cooled (ZFC) and Field-Cooled (FC) protocols. The samples were first cooled to the base temperature (2K) at zero magnetic field, after which 10 Oe field was applied and the magnetization was recorded. Subsequently, the sample was cooled in the applied field and the magnetization was measured again during the warming cycle under the same field conditions.

**Acknowledgement:**

AS acknowledge IIT Mandi for research facilities and DST India for Indo-Sweden bilateral grant (Grant No. DST/INT/SWD/VR/P-18/2019). PE acknowledges funding from the Swedish Government Strategic Research Area in Materials Science on Functional Materials at Linköping University (Faculty Grant SFO-Mat-LiU No. 2009 00971), the Knut and Alice Wallenberg foundation through the Wallenberg Academy Fellows program (KAW-2020.0196, P.E.), and the Swedish Research Council (VR) under Project Nos. 2021-03826 (P. E.), 2025-03680 (P. E.) and 2025-03760 (A. F.). A.S and SKH acknowledge Mauricio Sortica, Tandem Laboratory, Uppsala University for RBS and ERDA measurements. AM and SB thank the Department of Atomic Energy (DAE), Government of India and Chanakya post-doctoral fellowship, i-Hub quantum technology foundation (Sanction Order No. I-HUB/PDF/2022-23/04), for the financial support.

**Conflict of Interest:**

The authors declare no competing interests.

**Data Availability**

The data that support the findings of this study are available from the corresponding author upon reasonable request.

# <u>Supporting Information</u>

## Robust Superconductivity and High Upper Critical Fields in Epitaxial cubic W$_2$N Thin Films

Aditya Singh[1], Arnaud le Febvrier[2], Sanath Kumar Honnali[2], Abhisek Mishra[3], Grzegorz Greczynski[4], Subhankar Bedanta[3], Per Eklund[2, *], Ajay Soni[1, *]

[1]School of Physical Sciences, Indian Institute of Technology Mandi, Mandi 175005, Himachal Pradesh, India
[2]Department of Chemistry - Ångström Laboratory; Inorganic Chemistry, Uppsala University, Uppsala, 75105, Sweden
[3] Laboratory for Nanomagnetism and Magnetic Materials (LNMM), School of Physical Sciences, National Institute of Science Education and Research (NISER), An OCC of Homi Bhabha National Institute (HBNI), Jatni 752050
Odisha, India.
[4]Thin Film Physics Division, Department of Physics, Chemistry, and Biology (IFM), Linköping University, Linköping SE-581 83, Sweden
*Author to whom correspondence should be addressed: per.eklund@kemi.uu.se, ajay@iitmandi.ac.in

The supporting information has the additional details of characterization and experiments complementing to the main text.

### A. X-Ray diffraction of thin film and Thick Film

**Figure S1.** shows the XRD patterns of W$_2$N thin film (54 nm) and thick film (270 nm) deposited on Al$_2$O$_3$ substrate. The diffraction peaks for both samples show excellent phase matching, confirming identical crystal structures.

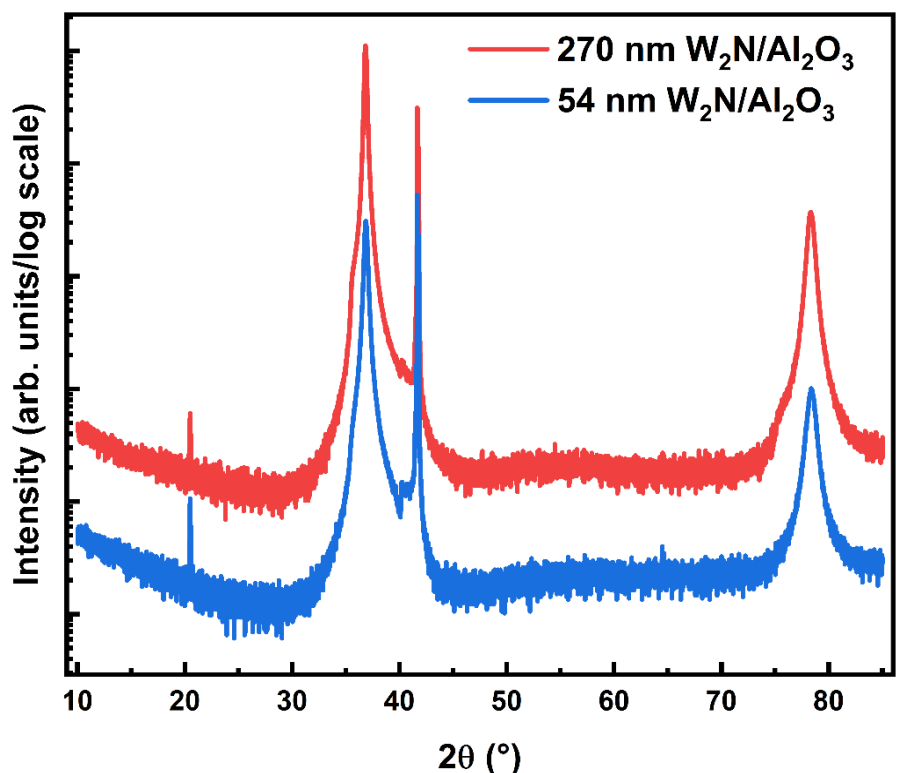

Figure S1: XRD patterns of thin film with XPS composition of W$_2$N$_{1.04}$ and the thick film W$_2$N/Al$_2$O$_3$

## B. Elemental Composition analysis using ERDA, RBS and XPS

The composition analysis of the thin films by XPS, RBS and ERDA are tabulated in table S1 for $W_2N/Al_2O_3$ and WN/MgO in addition to thinner film deposited on $Al_2O_3$ substrate. The table confirms that the deposited thin films are tungsten nitride ($W_2N$) with stoichiometries very close to the ideal composition. For the thick film grown on $Al_2O_3$ (WN/$Al_2O_3$) substrate, the extracted composition is $W_2N_{1.08}$, while for the film deposited on MgO substrate (WN/MgO) it is $W_2N_{1.14}$. The minor deviations from exact stoichiometry are typical in sputtered nitride thin films and remain within the expected range for stable $W_2N$ formation. Together, these results provide strong evidence that the deposited layers are indeed $W_2N$ thin films with near-ideal stoichiometry on both the substrates. A minor oxygen signal (~2.5 at.%) detected in the $W_2N/Al_2O_3$ film is near the detection limit of the analytical techniques and likely originates from surface oxidation. Overall, all three compositional analyses consistently confirm the formation of $W_2N$ with near-ideal stoichiometry on both substrates.

| Table S1: Composition analysis using ERDA, RBS and XPS | | | | | | |
|---|---|---|---|---|---|---|
| **Sample** | **XPS** | **ERDA and RBS** | | | | **Chemical Formula** |
| | N/W | N/W | At. %W | At. %N | At. %O | |
| **$W_2N$/MgO** | - | 0.57 | 63.8 | 36.2 | - | $W_2N_{1.14}$ |
| **270 nm $W_2N$/$Al_2O_3$** | - | 0.54 | 62.5 | 34.0 | 2.5 | $W_2N_{1.08}$ |
| **54 nm $W_2N$/$Al_2O_3$** | 0.52 | - | - | - | - | $W_2N_{1.04}$ |

## C. Pole Figure Analysis

The in-plane orientation of the films is demonstrated by the pole figure analysis of the 200-reflections of $W_2N$, observed at $2\theta \sim 43.04°$ for $W_2N/Al_2O_3$ and $\sim 42.94°$ for $W_2N/MgO$, as shown in Fig. S2(a-b). The six poles of $Al_2O_3$ $11\bar{2}3$, at $\Psi = 61.3°$ and six poles of c-$W_2N$ 200, at $\Psi = 56°$ in Fig S3(a) indicates epitaxial twin-domain growth, a common feature for cubic structures on both the substrates. In Fig S2 (b) three prominent poles are observed at $\Psi = 55.5°$, corresponding to MgO 200-reflections from the substrate alongside six poles from the $W_2N$ (200) planes, spaced every 60° in $\varphi$, with some overlapping the MgO poles. Both films epitaxially grow on c-plane sapphire and MgO (111) with no notably differences even in terms of twin domains which seems to be present on both substrates.

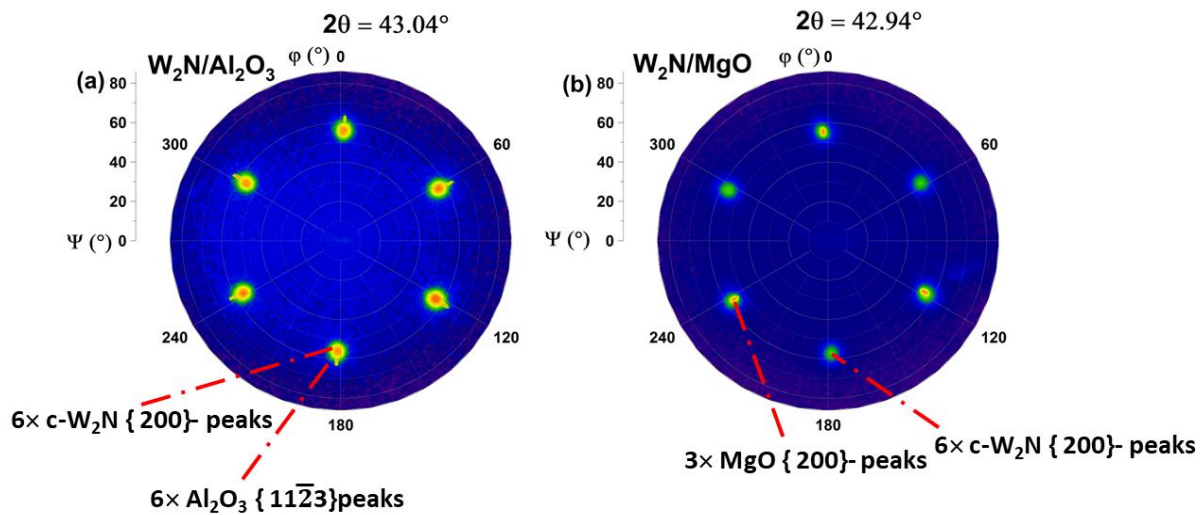

Fig S2: Pole Figure XRD for (a)W$_2$N/Al$_2$O$_3$ and (b) W$_2$N/MgO.

### D. Resistance Analysis using Small Polaron Hopping Model

The high temperature resistance is analysed with small polaron hopping (SPH) model which relates the temperature dependent resistivity $\rho(T)$ by the given expression $\rho(T) = \rho_0 T \exp\left(\frac{E_p}{k_b T}\right)$, where $E_p$ is the polaron activation energy. The experimental data exhibit a good fit to the SPH model (Fig. S3) indicating that the electrical transport mechanism is dominated by polaron hopping, arising from strong electron–phonon interactions that localize charge carriers and facilitate their migration through thermally activated hopping between lattice sites.

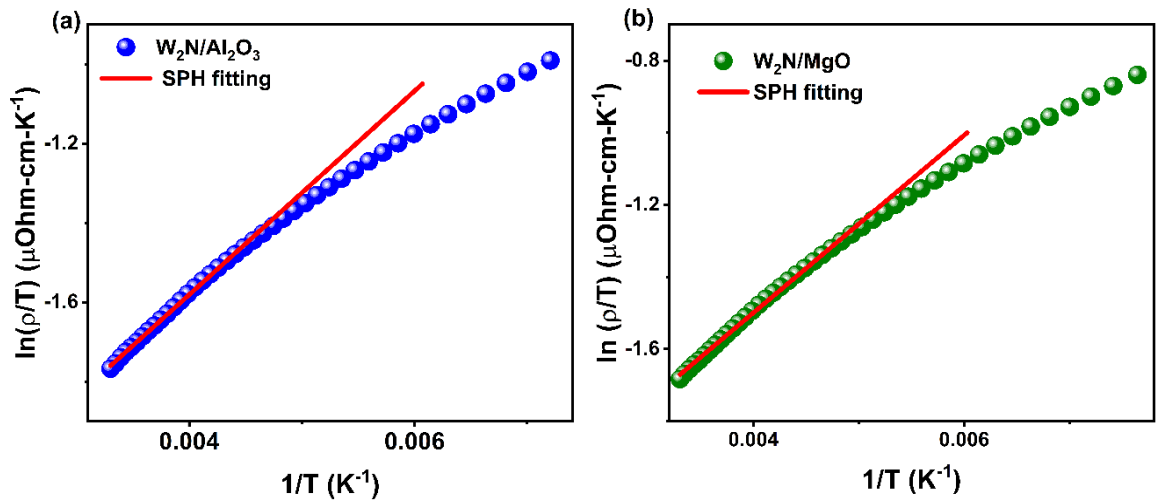

Fig. S3 Small Polaron hopping fitting of Resistance of (a) WN/Al$_2$O$_3$ and (b) WN/MgO